\documentstyle[preprint,eqsecnum,aps]{revtex}

\begin{document}

\draft
\title{Comment on the paper: "Water content and its effect on ultrasound  propagation in concrete- the possibility of NDE". Ultrasonic 38(2000) 546-552 by Etsuzo Ohdaira and Nobuyoshi Masuzawa}

\author{Manuel Rodriguez Gonzalez\thanks{mrg@mfc.uclv.edu.cu},Rolando Bonal Caceres\thanks{bonal@mfc.uclv.edu.cu}}

\address{Departamento de Fisica. Universidad Central de Las Villas. Santa Clara. CP: 54830 Villa Clara. Cuba}

\date{\today}

\maketitle

\begin{abstract}
The fundamental application of the ultrasonic pulse velocity method (UPV)  in the study of concrete, consist in the analysis and estimation of the mechanical properties (compressive strength). The precission of the estimation may be dependent on the incidence of various factors, one of which is the water content acquired by the sample of concrete. In Ref\cite{Etsuzo} a lineal dependence of the UPV in terms of the water content for different dosifications of concrete is proposed. This paper presents an analysis of  that dependence proving that an exponential model  describes better the behavior  of the velocity as function of water content.

\end {abstract}

\section{Introduction}
Concrete is one of the most important materials used is the development of a country's infrastructure. Each 
year, thousands of tons of concrete is used in the restoration  and rehabilitation of buildings. The development of the construction industry has helped in finding more effective and economic methods of inspection, which permit quicker and more reliable control. For this  reason, the application of non destructive test methods is becoming more widely used in the industry, among then the ultrasound Ref\cite{{Kathleen},{Ande},{Malhotra}}. Nevertheless, the effective application of the method in the study of concrete depends on the knowledge of a series of factors that affect  the pulse velocity as well as the mechanical parameters Ref\cite{Popovics 1}. One of these factors is the water content of  the concrete , in terms of which the ultrasonic pulse velocity can increase or decrease significantly , while the variations in the compressive strength remains insignifican Ref\cite{Manuel1}. The incidence of this factor can bring about imprecision in the estimation of the mechanical properties , specifically the strength of the concrete. The results obtained in Ref\cite {Etsuzo}are being analyzed in this paper and a model adequately describing the behavior of the velocity in terms of the water content is presented.

\section{Methods of measurement}
In the work Ref\cite {Etsuzo}, cylindrical samples of 200x100mm (diameter)are used for various dosifications. The specimen is situated on the scale to measure the variation of the water content according to weight lost ( until weight is constant). The transducer are attached  using the transmission method, which allows the measurement of the velocity in terms- of the water content.

\section{Discussion of the results.}

In Ref\cite {Etsuzo}, it is proposed that behaviour of the velocity as function of the concentration of water shows a lineal behavior, which is not particularly precise. If the linnear model is considered , it must be admitted that an increase in the concentration, in the absorption process, would always provoke an increase in velocity  which would not always  be possible from the physical point of view because the saturation process is  asymptotic, meaning that the velocity must reach an asymptotic maximum.  In figures 1,2,3, the non linnear tendency of the process is shown. From the calculation of the probability of the goodness of fit it is observed that in all cases , there are no significant  differences amoung velocity  for a significance level of 5 

It  is then necessary to find a model which adequately describes the phenomenon under study. In spite of the fact that a 3 rd order polynomial describes this process rather well from a statistical point of view, it lacks common sense when seen from a physical point of view. We have a phenomenon in which the extreme values have an asymptotic  behavior as a result of the drying process of the sample. As it is known, the increase of water in the material results in the increase in the velocity of  ultrasonic signal. So, if it is taken into account that the material has a saturation point which is reached asymptotically, then the velocity also has a maximum  value reached asymptotically. The former considerations can be written as follows:

Although the decrease in the concentration is proportional to the decrease in velocity, this decrease is greater while the velocity moves further away from a certain maximum value 

This affirmation can be expressed by the following differential  equation:

\begin{equation}
\frac{dV}{dC}=k(V_{0}-V)  \eqnum{1}
\end{equation}
And the solution is:

\begin{equation}
V=V_{0}-A_{0}e^{-kx}  \eqnum{2}
\end{equation}

Where: the integration constant A0 is considred negative in terms of the observations made the experiment.
v-velocity
V0, A0 parameters of the model
C-water content
K-parameter of the model

Adjusting for  each one the proportions , the following coefficients are obtained

\ \ \ \ \ \ \ \ \ \ \ \ \ \ \ \ \ \ \ \ \ \ \ \ \ \ \ \ \ \ \ 
\begin{tabular}{cccc}
Parameters & Test Piece A-1 & Test Piece B-1 & Test Piece C-1 \\ 
$V_{0}$ & $4487.58235$ & $4406.20072$ & $4234.08337$ \\ 
$A_{0}$ & $553.01462$ & $523.59085$ & $429.80387$ \\ 
$\frac{1}{k}$ & $2.88231$ & $2.20381$ & $3.07733$ \\ 
$R^{2}$ & $0.97$ & $0.97$ & $0.92$%
\end{tabular}

In all cases , the models converge rapidly.

From (2) it is observed that as concentration decreases, velocity decreases gradually and as C tends to zero , velocity assumes its minimum value, analogously , as C tends to infinite, velocity achieves its maximum value, describing a gradual process which corresponds with the capillary desorbption  and diffusion phenomenon, which depends on time and concentration  gradient. There are two stages involved in this drying process: initially capillarity, where the free water molecules rise to the pores´ surface    and evaporate into the surrounding air, and the diffusion period, which is responsible for the evaporation of water trough the fine pores. In order to know how capillarity varies  with time and concentration, it is necessary to solve the following equations for specific boundary condition of the desorbption process Ref\cite {{Nicos},{C.Hall 1},{C.Hall 2},{S.K},{WJ}}
  
\begin{equation}
\frac{\partial \theta }{\partial t}=\nabla D\nabla \theta   \eqnum{3}
\end{equation}

Where D-capillary  diffusivity
      t-time  
      $\theta$--- concentration of water.

Capillarity as well as diffusion, depend in great measure on the porosity of the medium. Therefore the coefficients are specific for each type of concrete.

Although the loss of water accompanied by contractions corresponds  to a cause and effect situation, their relationship is not a simple one. When concrete begins to dry up, the water lost first is the free water molecules contained in the capillaries , which practically produces no contraction Ref\cite {WJ}. For this reason , the first experimental point where a sudden variation of velocity occurs can be rejected.

Bearing in mind what has been said so far, an the existing dependence between velocity and the acoustic characteristic of the medium , it can then be understand why velocity varies whit the sample water content Ref\cite {{Popovics2},{Popovics3},{Popovics4},{Popovics5},{Popovics6}}. There is a significant difference between the acoustic impedance of the air which exists initially in the superficial   pores and that of the water which is absorbed in the course of time , in other words , if the pores are filled with air , the medium is denser , which favors the propagation's of the high- frequency components which are more sensitive to these change. As  observed in Ref\cite {Etsuzo}, on obtaining the frequency spectrum, the value  of the fundamental component decreases(attenuate) as the concentration of water decreases.
However, the acoustic impedance of water and air are small in comparison with impedance of concrete. Why is it then that this variations in velocity occurs? It is logical that the wave is propagated preferably through the solid matrix, but if the number of pores is very high, then velocity is affected by the presence of pores. Besides, water penetrates troungh a distance of 3-4 cm in the interior of the sample. This implies that all the area near the surface is saturated and its dimensions are  of the order of the wavelength of the fundamental component , meaning that for velocities above 4000 m/s and frequencies of the order of 55- 100 KHz , the wavelength of the fundamental component is about 4-6 cm, the entire pulse front being affected by the presence of water.

It is important to demonstrate that the relative error of  estimated velocity can be calculated from the model for any given concentration using the following expression:

 \begin{equation}
\delta V=0.0108895+\frac{-0.96646-0.0092429C}{-0.123232+e^{0.3469446C}} 
\eqnum{4}
\end{equation}

where: delta V-------- relative error of velocity
       C ------concentration.

This error, as observed in Figura 4, it is approached to one percent in the low water content zone and diminishes with the increase of concentration as a consequence of the asymptotic tendency of the saturation process.

It is known that the fundamental objective of the application of ultrasound in the study of concrete is to correlate velocity witch compressive strength Ref\cite{{B.E},{DiMaio1},{DiMaio2},{Ferreira},{Ibrahim}}, however when the humidity condition changes as has been analyzed, the behavior of the strength is not analogous to that of velocity, it practically maintains an almost constant value, as shown in Ref\cite {Manuel1} for this reason, it is very important to keep in mind that high velocity do not always imply that values of strength will also be high, so mistakes can be made in the estimation of strength. In Figure 5 it is shown the tendency of estimated strength using the expression (5) Ref\cite {Wee}, employing the value of experimental velocity obtained in Ref\cite {Etsuzo}, the velocity estimated using the linnear model proposed in Ref\cite {Etsuzo} and velocity estimated through the exponential model proposed in this work. In all cases velocity increases with water content. It is observed that estimated strength also augments with water content, which is incorrect. That's why to estimate the strength from velocity it is necessary to know moisture percent of the sample.

\begin{equation}
R=aV-b  \eqnum{5}
\end{equation}

where: R- compressive strength (Mpa)
       a,b-parameters of the model
       V-velocity

In this case we employ a=0.0447 y b= 149.67 Ref\cite {Manuel 2}

\section{conclusions}
 It is proven physically and statistically that the model which best describes the behaviour of velocity with respect  to water concentration is the decreasing exponential model, from which the compressive strength can be estimated with a good confidence.

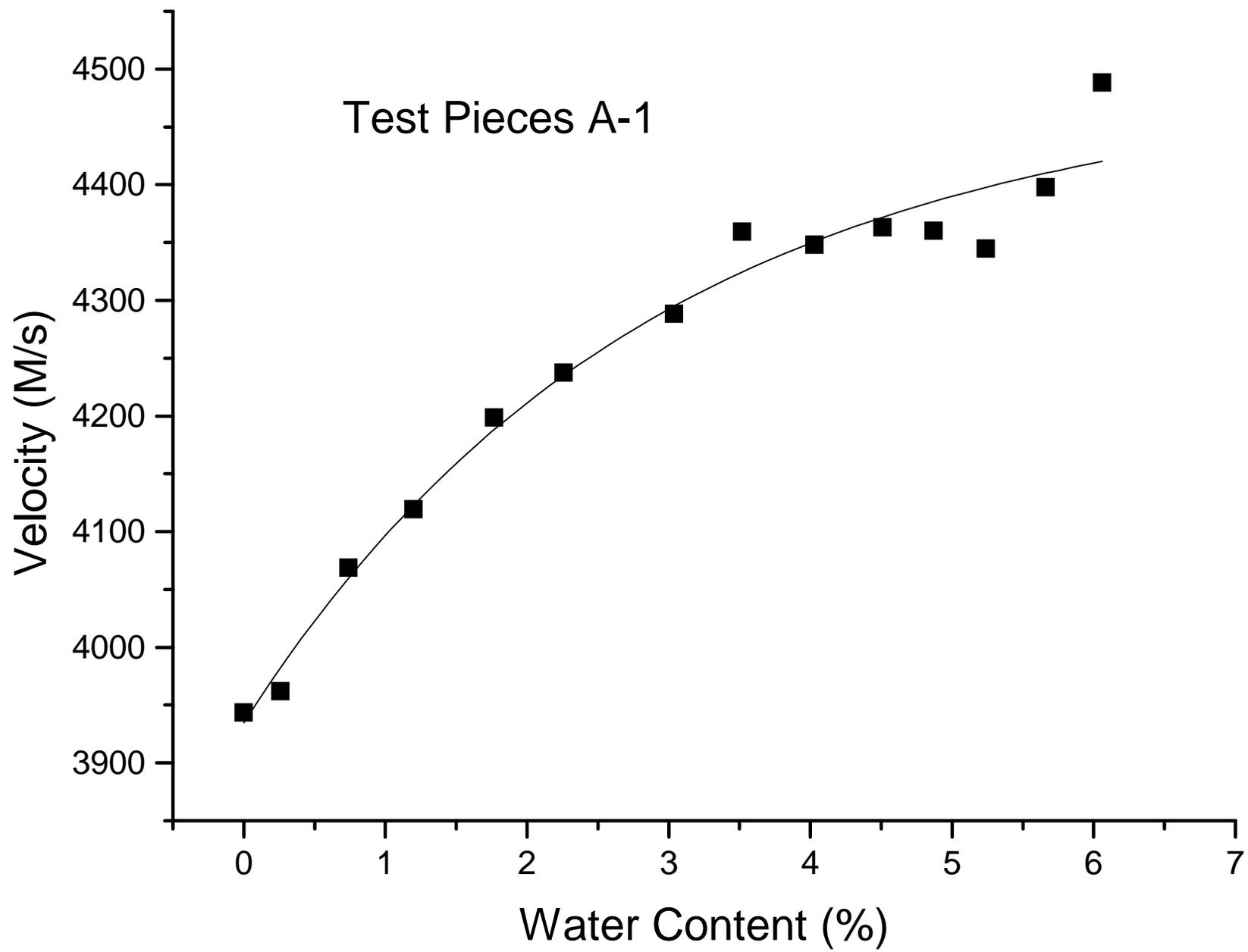
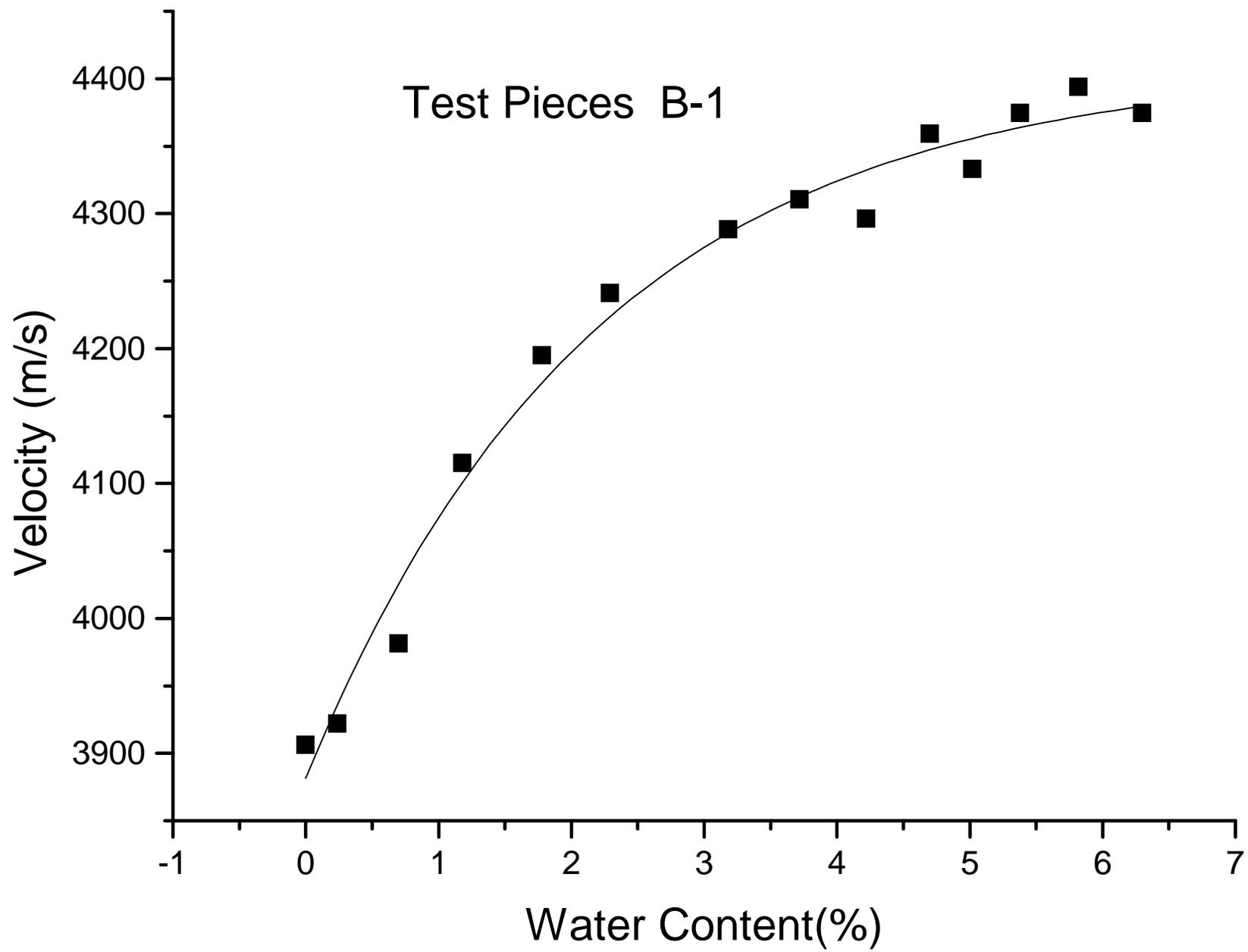
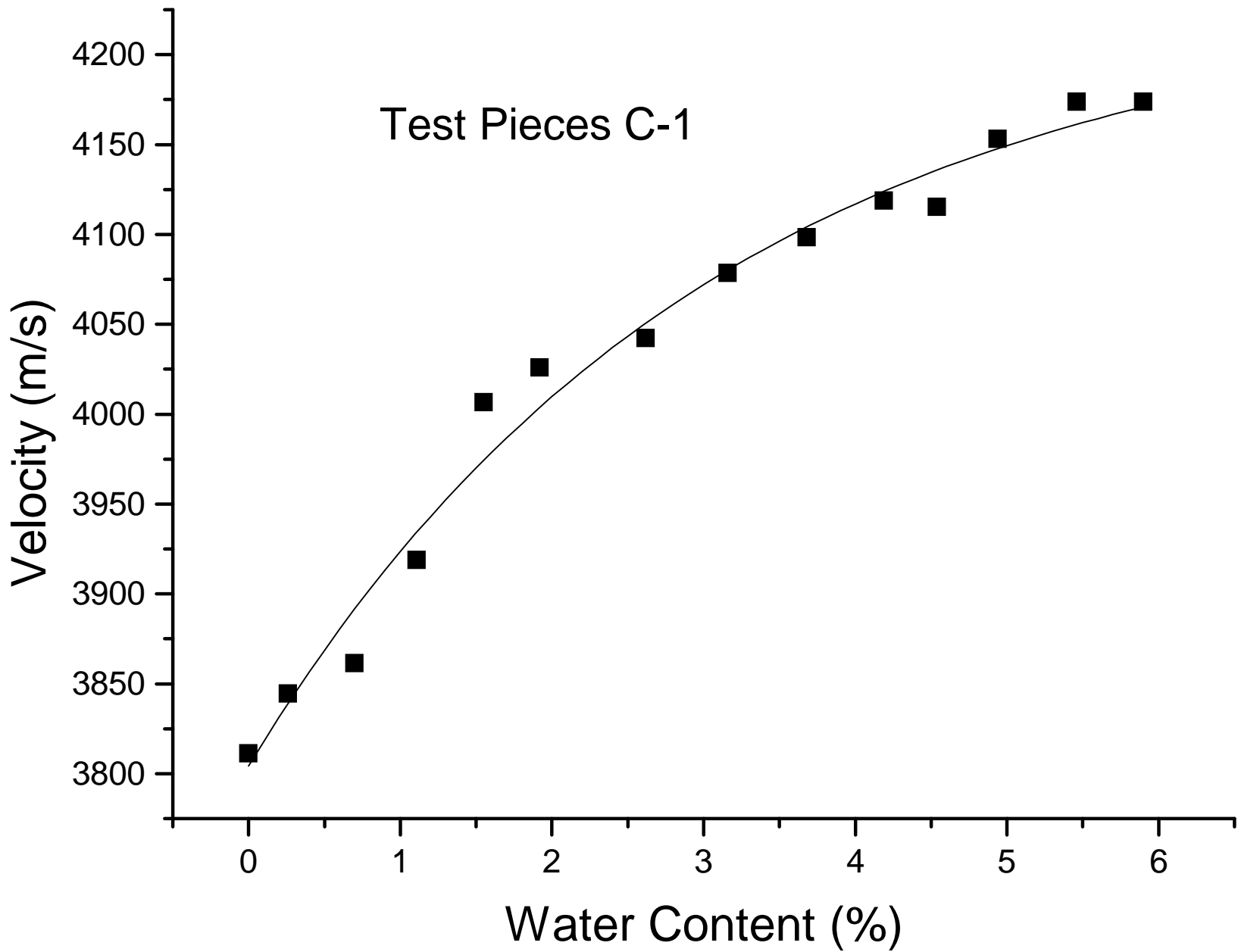
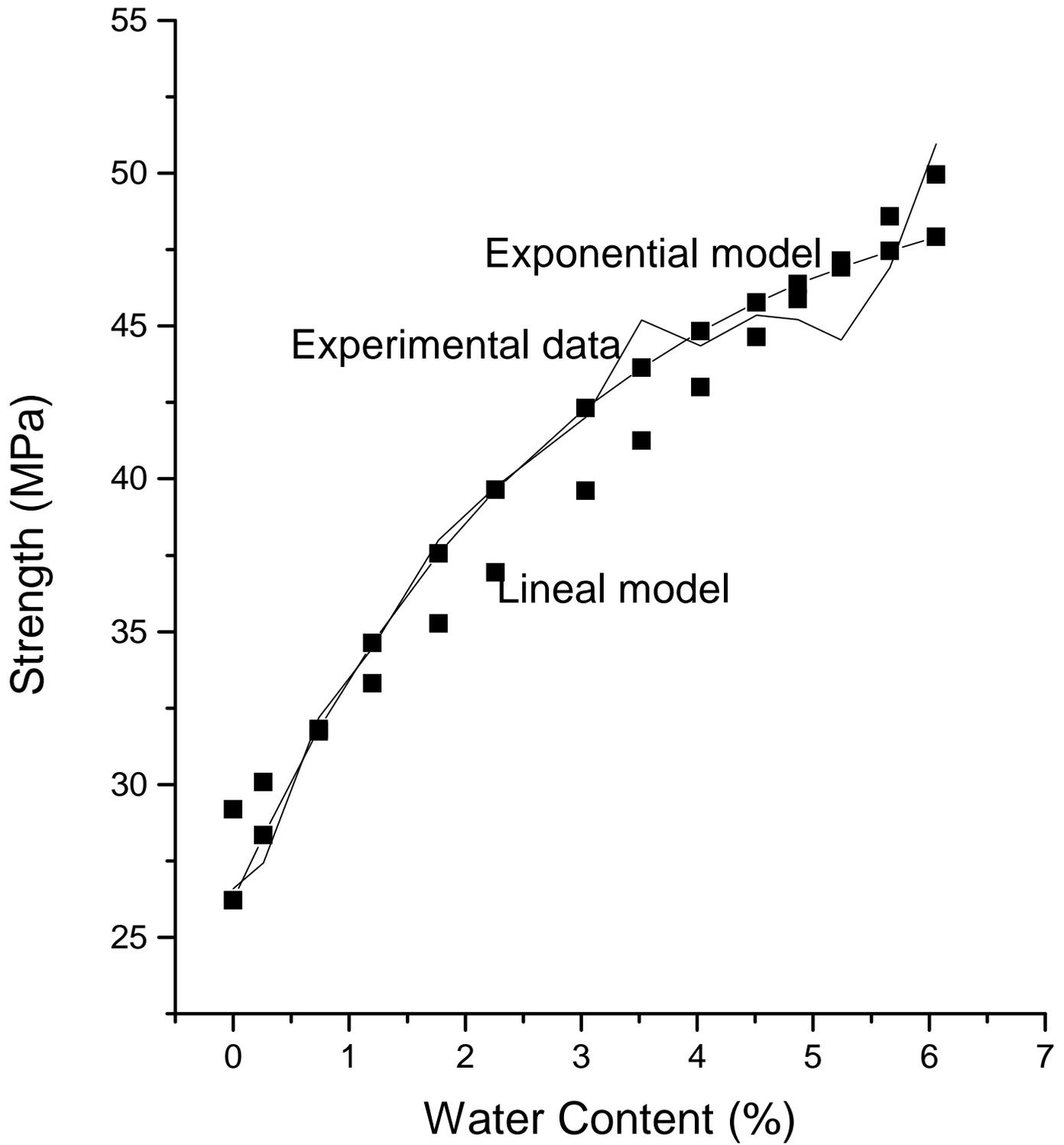
\end{document}